\documentclass[aps,prb,preprint,groupedaddress,showpacs,amsmath,amssymb]{revtex4}

\usepackage{graphicx}% Include figure files
\usepackage{dcolumn}% Align table columns on decimal point
\usepackage{bm}% bold math

\begin{document}

%Title of paper
\title{Effect of pressure on transport and magnetotransport properties in CaFe$_2$As$_2$ single crystals}

\author{M. S. Torikachvili}
\affiliation{Department of Physics, San Diego State University, San Diego, CA 92182-1233, USA}
\author{S. L. Bud'ko}
\author{N. Ni}
\author{P. C. Canfield}
\affiliation{Ames Laboratory US DOE and Department of Physics and Astronomy, Iowa State University, Ames, Iowa 50011, USA}
\author{S. T. Hannahs}
\affiliation{National High Magnetic Field Laboratory, 1800 East Paul Dirac Drive, Tallahassee, FL 32310, USA}

\date{\today}

\begin{abstract}

The effects of pressure generated in a liquid medium, clamp, pressure cell on the in-plane and $c$-axis resistance, temperature-dependent Hall coefficient and low temperature, magnetoresistance in CaFe$_2$As$_2$ are presented. The $T - P$ phase diagram, including the observation of a complete superconducting transition in resistivity, delineated in earlier studies is found to be highly reproducible. The Hall resistivity and low temperature magnetoresistance are sensitive to different states/phases observed in CaFe$_2$As$_2$. Auxiliary measurements under uniaxial, $c$-axis, pressure are in general agreement with the liquid medium clamp cell results with some difference in critical pressure values and pressure derivatives. The data may be viewed as supporting the potential importance of non-hydrostatic components of pressure in inducing superconductivity in CaFe$_2$As$_2$.

\end{abstract}

% insert suggested PACS numbers in braces on next line
% insert suggested keywords - APS authors don't need to do this
%\keywords{}

%\maketitle must follow title, authors, abstract, \pacs, and \keywords
\maketitle

\section{Introduction}
In the recent hemorrhage of exciting experimental and theoretical results related to novel superconductivity in Fe-As containing materials, studies of CaFe$_2$As$_2$ take a rather special place. This compound was synthesized very recently, in form of large single crystals. \cite{nin08a,ron08a,wug08a} At ambient pressure, at $\sim 170$ K, a first order structural (from high temperature tetragonal to low temperature orthorhombic) phase transition (with a few degrees temperature hysteresis) was observed. \cite{nin08a} The structural phase transition was found to be coincident with an antiferromagnetic ordering of Fe moments in the $ab$ plane, \cite{gol08a} bearing similarity to BaFe$_2$As$_2$ and SrFe$_2$As$_2$. \cite{rot08a,nin08b,mit08a,kre08a,yan08a,zha08a} The anisotropic three-dimensional magnetism in CaFe$_2$As$_2$ was further studied by inelastic neutron scattering and band structure calculations. \cite{mcq08a} Even more striking was  the observation of pressure-induced superconductivity with $T_c \le 12$ K observed in CaFe$_2$As$_2$ \cite{tor08a,par08a} at very moderate, $P \sim 5$ kbar pressures. Following the work on CaFe$_2$As$_2$, pressure-induced superconductivity was also found in BaFe$_2$As$_2$ and SrFe$_2$As$_2$, \cite{ali08a,iga08a,kot08a,man09a,col09a,kot09a} albeit at significantly higher pressures. Neutron scattering studies of the magnetic and structural properties of CaFe$_2$As$_2$ under hydrostatic (He gas and liquid media cell) pressure \cite{kre08b} elucidated the complex $P - T$ phase diagram that has been roughly outlined as a result of electrical transport measurements under pressure. \cite{tor08a} Three different magnetic/crystallographic  phases were identified below room temperature under pressure up to $\sim 6.5$ kbar: in addition to non-magnetically ordered, tetragonal (T) and antiferromagnetic orthorhombic (O) phases observed at ambient pressure, \cite{gol08a} a non-magnetic, "collapsed tetragonal" (cT) phase occured in CaFe$_2$As$_2$ at low temperatures for pressures above $\sim 3$ kbar. This cT phase is stabilized by further increases in pressure, rising to room temperature by $\sim 17$ kbar. \cite{tor08a,kre08b} Band structure calculations \cite{kre08b} confirmed the non-magnetic character of the cT phase. Upon comparison with the phase diagram from transport measurements under pressure, \cite{tor08a} a conjecture that superconductivity dome appears in the non-magnetic cT phase was proposed. \cite{kre08b}. This picture appears to be in a contradiction with the interpretation of the $\mu$SR experiments under pressure (using Daphne oil 7373 as a pressure medium) \cite{gok08a} in which a co-existence of superconductivity with a magnetic order in a partial volume fraction was inferred. On the other hand, recent transport measurements under pressure (using silicone fluid as a pressure medium), \cite{lee08a} although confirming most of the results presented in Ref. \onlinecite{tor08a}, alleged that the extent of existence of the cT phase is limited by $\sim 8$ kbar below $\sim 150$ K, with a new, unidentified, phase located at higher pressures. (Later, after the appearance of several He pressure cell results, \cite{kre08b,yuw08a,gol08b} the interpretation of the same set of data was significantly altered. \cite{lee08b}) In addition to aforementioned apparent dissimilarities in the interpretation of the experimental data, disparate views on the evolution of properties and ground states under pressure exist in band structural results as well. \cite{kre08b,yil08a,sam08a} Finally, recent resistivity and magnetic susceptibility measurements using helium as a pressure medium \cite{yuw08a} have confirmed, but with {\it very sharp} features in resistivity, the phase lines associated with tetragonal to orthorhombic/antiferromagnetic as well as with tetragonal to the collapsed tetragonal phase transitions. However no superconductivity was observed in either measurement, leading to the conjecture \cite{yuw08a,gol08b} that a non-hydrostatic component of pressure is required to induce superconductivity in CaFe$_2$As$_2$.

The T to cT structural transition results in significant, anisotropic, chance of the lattice parameters. \cite{kre08b} For CaFe$_2$As$_2$ in liquid media pressure cells in the 5 - 10 kbar pressure range at the temperatures corresponding to T - cT transition the media is solid (polycrystalline or glass - like), so cooling through the structural transition may rather be thought of as a {\it constant volume} not {\it constant pressure} process. \cite{yuw08a} It is clear then that in liquid media cells the CaFe$_2$As$_2$ sample is subjected to non-hydrostatic stresses on cooling through T - cT transition and the effect is magnified by large lattice parameters/volume changes of the sample at this phase line. This results in multi-crystallographic phase at low temperatures \cite{gol08b} and the superconducting dome observed in liquid media pressure cells experiments, most probably originates in one or another stressed crystallographic phase.  A comprehensive overview of the physical properties of CaFe$_2$As$_2$ at ambient and high pressure is presented elsewhere. \cite{can09a}

Given that He cells used have limited pressure ranges (generally below 7 - 10 kbar) and bearing in mind the somewhat limited and possibly contradictory results on CaFe$_2$As$_2$ under pressure, in this work we present a detailed study of magnetotransport in this material under pressure (in a liquid medium, clamp pressure cell) that is extending our previous results, \cite{tor08a} supplemented by the data on the effect of uniaxial, along the $c$-axis, pressure on in-plane resistance in this material. The main goal of this study is to further explore the properties of CaFe$_2$As$_2$ under pressure generated in a pressure cell of a commonly used clamp type with a liquid medium so as to probe the superconducting state and conditions necessary to stabilize it.

\section{Experimental methods}

\begin{figure}
\begin{center}
\includegraphics[angle=0,width=120mm]{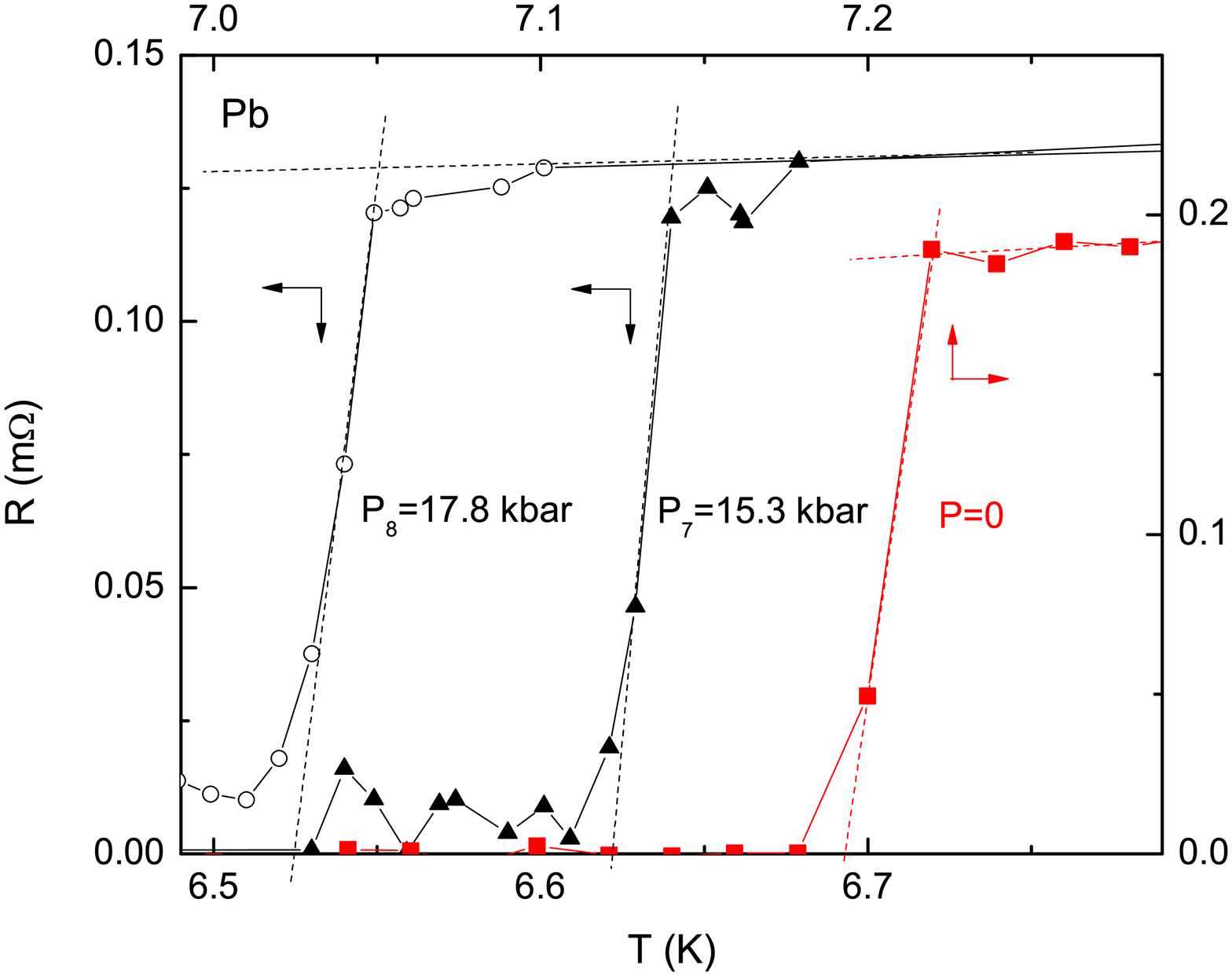}
\end{center}
\caption{(Color online) Example of resistive transitions in Pb, used as a low temperature manometer, for two pressures above 15 kbar (left and bottom axes) and at $P = 0$ (right and top axes). Dashed lines are guides for the eye.}\label{F1}
\end{figure}

Single crystals of CaFe$_2$As$_2$ were grown out of Sn flux (as discussed in Ref. \onlinecite{nin08a} in more detail) using conventional high temperature, solution growth techniques \cite{can92a}. Electrical transport and magnetotransport under pressure were measured for pressures below $\sim 20$ kbar generated in a Teflon cup filled with Fluorinert FC-75, with the cup inserted into a non-magnetic, piston-cylinder-type Be-Cu pressure cell with a core made of NiCrAl (40-KhNYu-VI) alloy. Pressure at low temperatures was determined by monitoring the $T_c$ of pure lead. \cite{eil81a}. The pressure generation and medium was the same as that used in our previous publication. \cite{tor08a} For consistency, low temperature pressure values will be used throughout the text. The uncertainty in pressure ($\sim 1$ kbar for higher temperature transitions) does not significantly effect our conclusions. The width of the superconducting transition in lead (together with the established $\Delta T_c(P)$ behavior \cite{eil81a} and bearing in mind smaller size of the sample in comparison to the Pb-manometer) can be used to evaluate the pressure gradients in the cell at low temperatures. (Here we ignore the fact that the Pb - transition has a finite, and very similar, width at ambient pressure.) Fig. \ref{F1} gives an estimate of $\Delta T_c = 0.02 - 0.03$ K, that could correspond to $\Delta P = 0.5 - 0.8$ kbar (or $< 5\%$ at the higher end of our pressure range). This, though, is absolutely an upper limit to $\Delta P$ since a similar $\Delta T_c$ is found for the Pb manometer at ambient pressure as well. This estimate however does not address the very peculiar situation \cite{yuw08a,can09a} caused by the cooling of the CaFe$_2$As$_2$ sample through a structural, tetragonal to collapsed tetragonal, phase transition with a large, {\it anisotropic}, change of the lattice parameters \cite{kre08b,can09a}, that almost invariably should cause an additional, non-hydrostatic, component of pressure, while the sample is constricted to the volume it occupied when the media froze.

The temperature and magnetic field environment for the pressure cell were provided by a Quantum Design Physical Property Measurement System (PPMS) instrument. The temperature of the sample was determined by an additional Cernox sensor attached to the body of the pressure cell. The cooling/warming rates were below 0.5 K/min. The resulting temperature lag between the Cernox on the body of the cell and the sample was $< 0.5$ K at high temperatures and 0.1 K or less below $\sim 70$ K. For one of the samples ($I \| (ab), H \|c)$) below $\sim 10$ K the resistivity was measured in 250 Oe field to suppress the superconductivity of traces of elemental Sn (residual flux).

Resistivity measurements under pressure in this work were performed for $H \| c$ orientation of the samples. The accuracy of the samples' orientation with respect to the applied field was $\sim 5^\circ$. Pt wires and Epo-Tek H20E silver epoxy were used to make contact to the samples. For the $I \| (ab)$ sample the contacts were positioned in a standard linear geometry. For (semi-quantitative) $I \| c$ measurements both current (larger) and voltage (smaller) contacts were positioned on the opposite, parallel, $(ab)$ faces of the sample. In this case the resulting resistivity has an admixture of both, $\rho_{ab}$ and $\rho_c$ components, however, presumably, with a large contribution of the latter (c.f. Ref. \onlinecite{son03a}). Hall measurements on CaFe$_2$As$_2$ under pressure were performed with current flowing in the $ab$ plane approximately parallel to the $a$-axis and field parallel to the $c$-axis. To eliminate the effect of a misalignment of the voltage contacts, the Hall measurements were taken for two opposite directions of the applied field, $H$ and $-H$, and the odd component, $[\rho_H(H) - \rho_H(-H)]/2$, was taken as the Hall resistivity. $\pm 90$ kOe magnetic fields were used for these measurements.

Additional field-dependent electrical transport data under pressure for CaFe$_2$As$_2$  single crystal were collected using the 310 kOe resistive magnet in the National High Magnetic Field Laboratory (NHMFL) in Tallahassee, FL with a $^3$He refrigerator insert that allowed for accommodation of the pressure cell. For these measurements electrical current was flowing in the $ab$ plane and the magnetic field was applied along the $c$-axis.

Resistance under uniaxial (along the $c$-axis) stress was measured in a home-made stress cell attached to a standard PPMS puck. A rectangular, uniformly thick, sample was uniaxially compressed between two phosphor bronze discs, with the surfaces in contact with the sample being electrically insulated. Thin, annealed, platinum wires were used to make the electrical contacts to the sample in a standard 4-probe geometry. The parts of the wires to be in contact with the sample were flattened before the experiment. (Mechanical contact under finite applied stress was enough to give contact resistances $\sim 1~\Omega$ or less.) The body of the cell was made of phosphor bronze, the stress was applied via a NiCrAl spring, that was calibrated (force vs. compression) at room temperature using an Omega, LCMDK-1KN, load cell. There is an uncertainty in the temperature dependence of stress caused by difference in thermal expansions between phosphor bronze and the NiCrAl alloy and, more importantly, the change in elastic properties of the NiCrAl alloy on cooling. We are not aware of temperature dependent elastic properties data for the NiCrAl alloy, however for many steels and alloys the change in elastic constants from room temperature to liquid helium temperature is rather small, less than 20\% [\onlinecite{led}], so in the following we will use the room temperature estimates for uniaxial pressure, $p_c = F/S$, where $S$ is the sample area. Given these approximations, the uniaxial stress is known (and controlled) at a semi-quantitative level.

In the following (unless stated otherwise) the onset criterion is used to infer the superconducting transition temperature and the extremum (minimum) of the $d \rho/d T$ derivative to infer the structural/magnetic transitions temperatures.

\section{Results and Discussion}

\begin{figure}
\begin{center}
\includegraphics[angle=0,width=80mm]{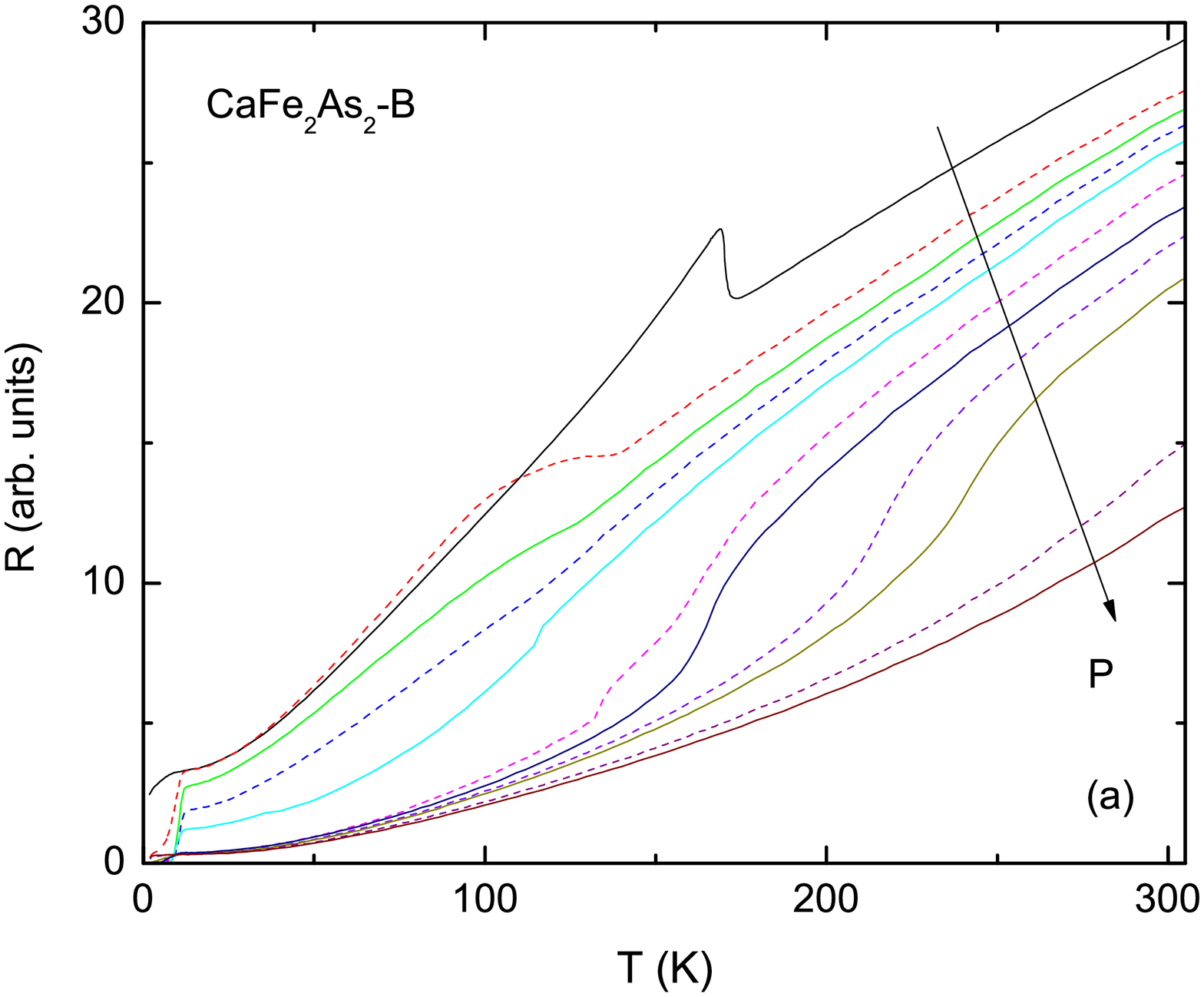}
\includegraphics[angle=0,width=80mm]{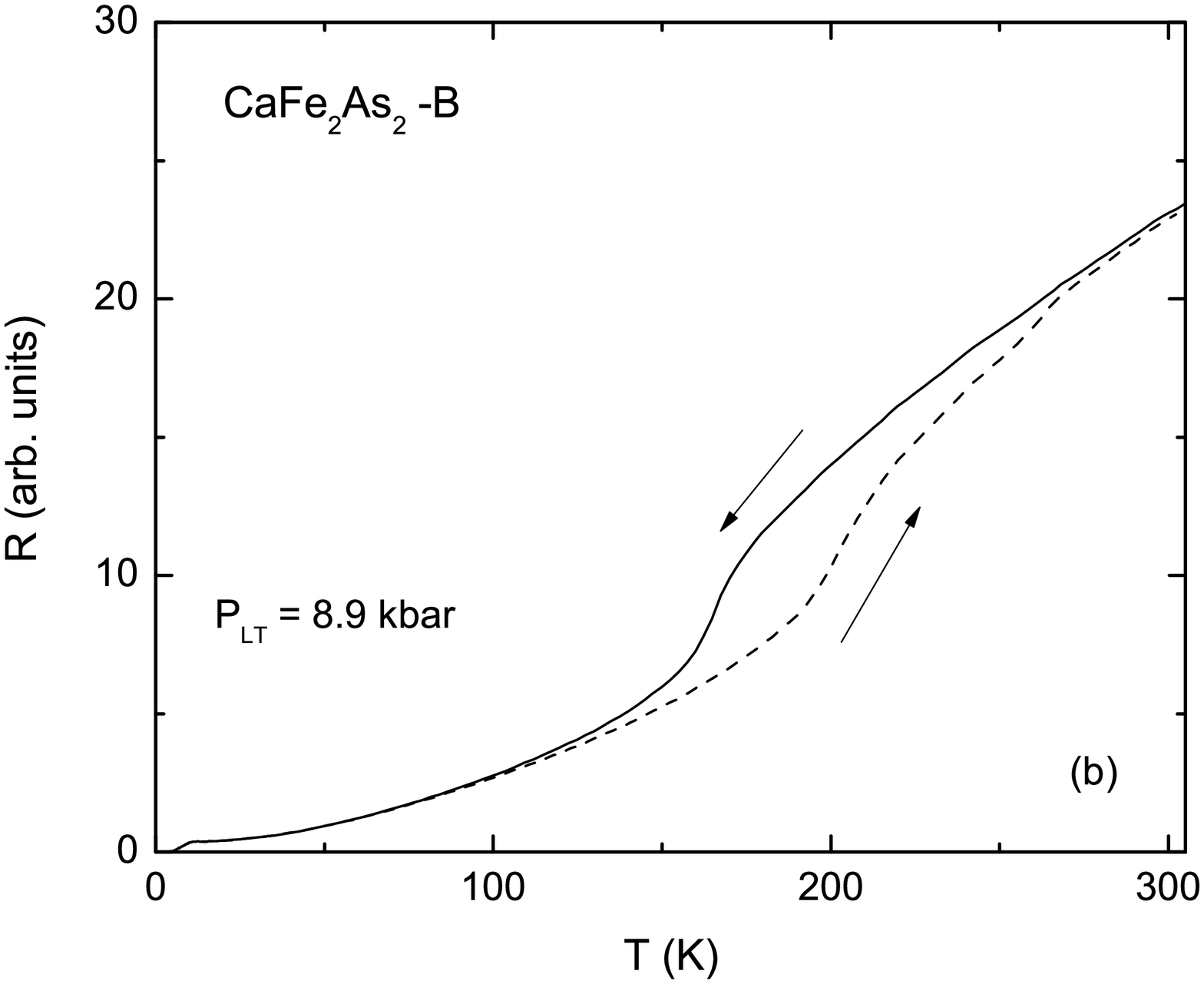}
\includegraphics[angle=0,width=80mm]{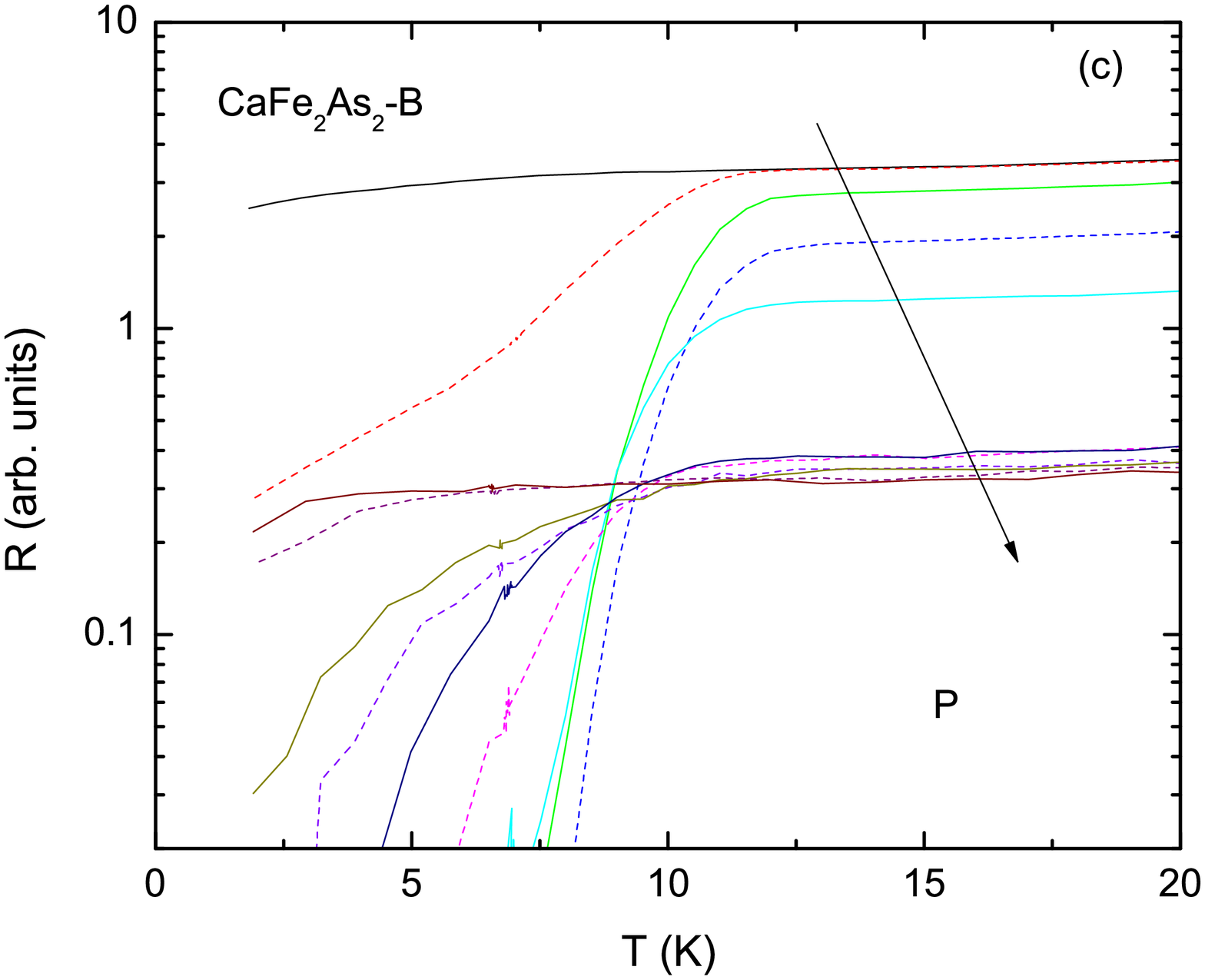}
\end{center}
\caption{(Color online) (a) Temperature-dependent in-plane resistance of CaFe$_2$As$_2$ at different applied pressure (0, 2.7, 4.3, 4.9, 6.4, 8.4, 8.9, 11.0, 12.1, 15.3, and 17.8 kbar, arrow is pointing to the direction of increase of pressure). Data taken on cooling are shown. (b) Data for $P = 8.9$ kbar taken on cooling and warming. Panel (c) - low temperature part of the data in the panel (a).}\label{F2}
\end{figure}

Fig. \ref{F2} presents the temperature-dependent resistance curves, $R(T)$, taken with current in the $(ab)$ crystallographic plane, from sample B of CaFe$_2$As$_2$ (we denote the sample used in Ref. \onlinecite{tor08a} as sample A). This set of data is in a very good agreement with our previous results \cite{tor08a} as well as  with the results (but not the interpretation) of Lee et al. \cite{lee08a}  The features associated with the T - O and T - cT phase transitions \cite{tor08a,kre08b,lee08a} are clearly seen in these data, including the large, $\sim 30$ K, hysteresis at T - cT phase line (Fig. \ref{F2}(b)) described in Ref. \onlinecite{tor08a}. A complete superconducting transition in resistance is observed for $4.3 \ge P \ge 11$ kbar, with noticeable broadening of the transition above 6.4 kbar (Fig. \ref{F2}(c)). In addition to superconductivity being stabilized at intermediate pressures Fig. \ref{F2}(c) shows the very non-monotonic drop of the residual resistivity with increasing pressure that has beed associated with the superconducting region. \cite{tor08a}

\begin{figure}
\begin{center}
\includegraphics[angle=0,width=120mm]{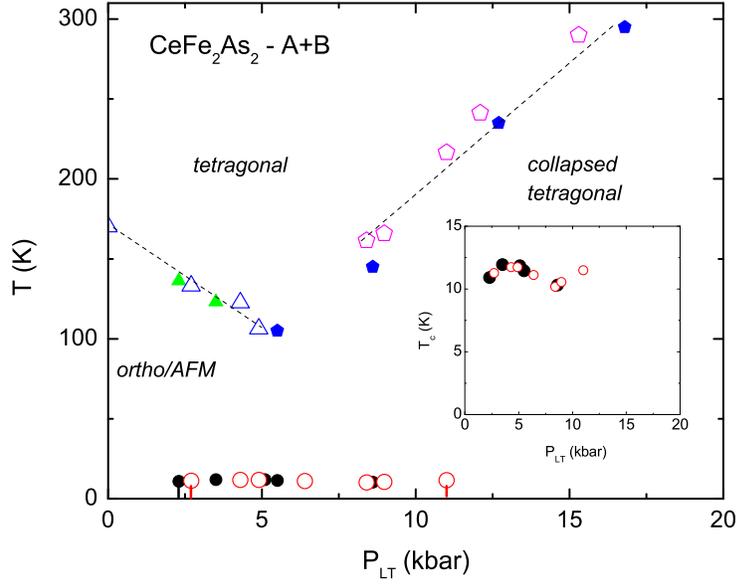}
\end{center}
\caption{(Color online) Combined phase diagram of CaFe$_2$As$_2$ under pressure. Open symbols - this work (sample B), filled symbols - previous data \cite{tor08a} (sample A). Circles - onset of superconductivity, triangles - structural/AFM phase transition, pentagons - T - cT phase transitions. Inset: enlarged $T_c(P)$ part of the phase diagram. Transition temperatures taken from $R(T)$ on cooling are shown. Dashed lines are guides for the eye.}\label{F3}
\end{figure}

\begin{figure}
\begin{center}
\includegraphics[angle=0,width=120mm]{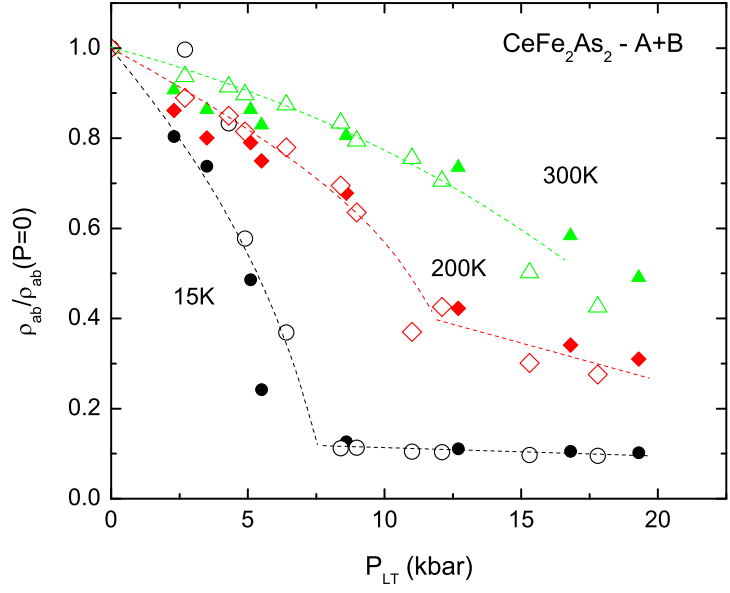}
\end{center}
\caption{(Color online) Normalized to the values at $P = 0$ room temperature in-plane resistivity, $\rho (300K)$, resistivity at 200 K, $\rho (200K)$, and low temperature - normal state in-plane resistivity, $\rho (15K)$ for two samples of CaFe$_2$As$_2$ under pressure. Open symbols - this work (sample B), solid symbols - previous data \cite{tor08a} (sample A). Dashed lines are guide to the eye.}\label{F4}
\end{figure}

The agreement between this work and our previous results is clearly seen in the combined phase diagram, Fig. \ref{F3} and in the pressure dependencies of the room temperature and low temperature - normal state resistivities, Fig. \ref{F4}. The dramatic reduction in $\rho_{ab}$ found at 15 and 200 K is associated with the greatly reduced resistivity in the cT phase. \cite{tor08a,yuw08a} At 300 K the transition takes place at the far edge of our pressure range \cite{tor08a,gol08b} and is hard to detect. It is worth noting the sizable pressure  dependence of the T - phase resistivity above the cT transition. This is not unique to CaFe$_2$As$2$, and is also large in BaFe$_2$As$2$ and SrFe$_2$As$2$. \cite{col09a}

\begin{figure}
\begin{center}
\includegraphics[angle=0,width=90mm]{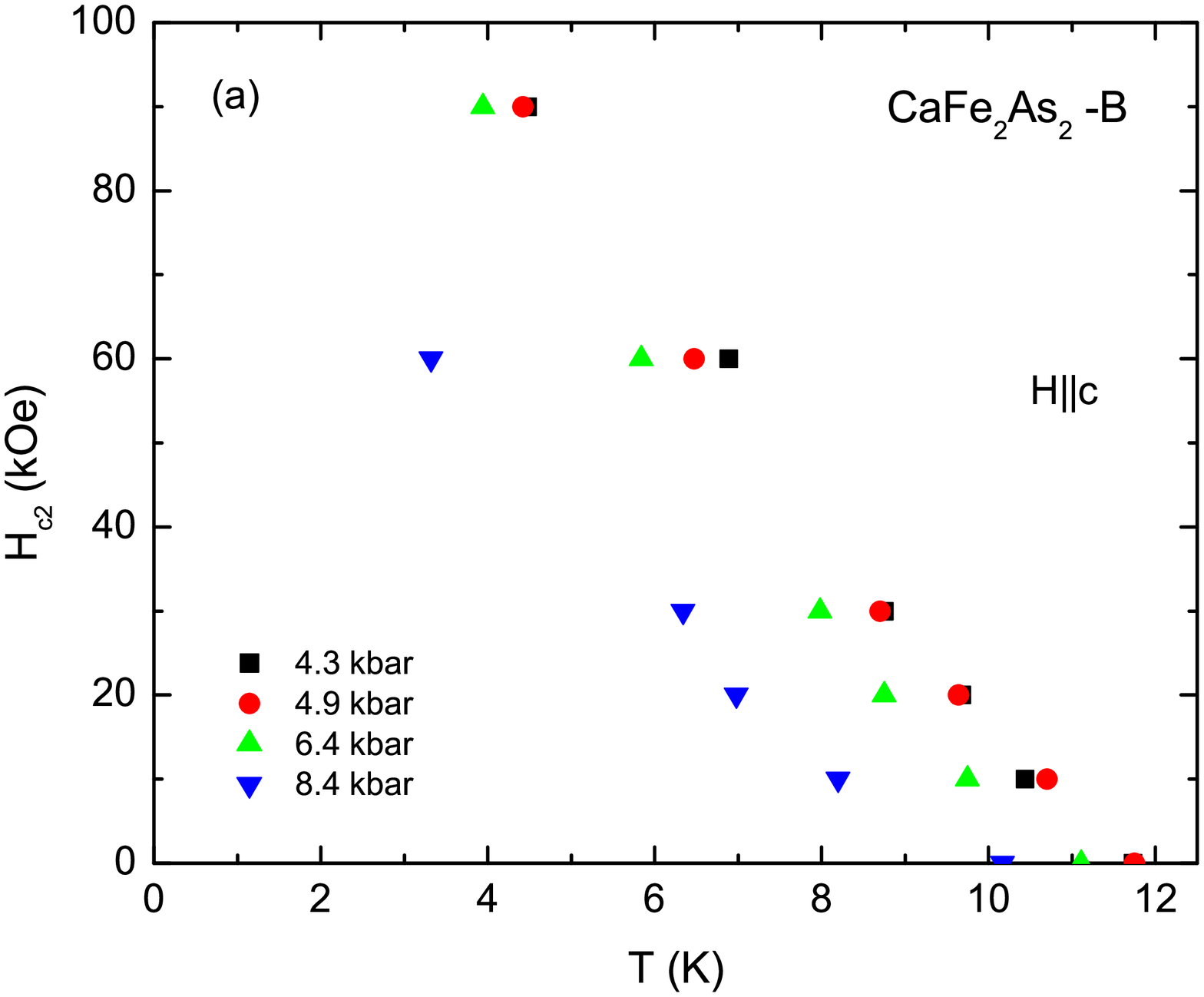}
\includegraphics[angle=0,width=90mm]{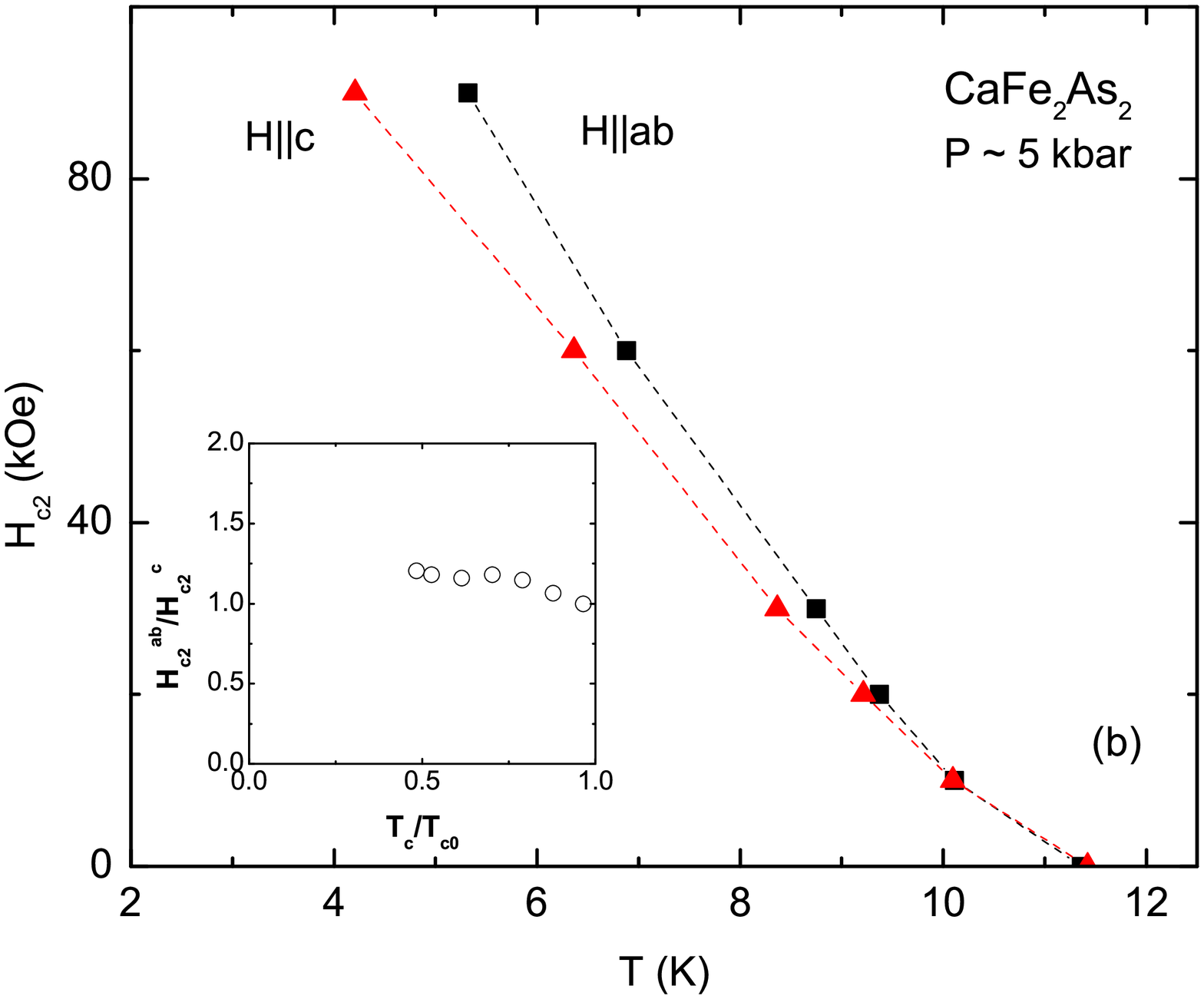}
\end{center}
\caption{(Color online) (a) Upper critical field, $H_{c2}(T)$, ($H \| c$) of CaFe$_2$As$_2$, determined from the onset of resistive transitions, at several values of pressure. (b) Anisotropic $H_{c2}(T)$ for $P \sim 5$ kbar: $P = 5.5$ kbar for $H \| ab$ and average between 4.3 and 6.4 kbar for $H \| c$ are shown. Inset: estimate of the temperature-dependent anisotropy of the upper critical field in the pressure - induced superconducting phase in CaFe$_2$As$_2$ at $P \sim 5$ kbar.}\label{F5}
\end{figure}

Sample B was placed in the pressure cell so that $H \| c$ (as opposed to $H \| ab$ in our previous work \cite{tor08a}). The $H_{c2}(T)$ curves determined from the onsets of resistive transitions for several pressures are shown in Fig. \ref{F5}(a). All of the curves have some upward  curvature for low fields. A rough extrapolation suggests that the $H_{c2}(0)$ values range (for different pressures and models) between 70 kOe and 140 kOe.

To evaluate the anisotropy of $H_{c2}$ in the pressure-induced superconducting phase of CaFe$_2$As$_2$, we combine the data sets for the two orientations of the applied field (obtained on two different samples) in Fig. \ref{F5}(b). Since these data were obtained for two different samples in two different pressure runs, exactly the same pressures and $T_c(H = 0)$ were not achieved. For a rough estimate of the anisotropy we compare data for  $P = 5.5$ kbar for $H \| ab$ and average  between 4.3 and 6.4 kbar ($T_c(H) = [T_c^{4.3kbar}(H) + T_c^{6.4kbar}(H)]/2$) for $H \| c$. (This gives a zero field $T_c$ closest to the 5.5 kbar data and allows for clearest estimate of anisotropy over widest field range.) The $H_{c2}$ anisotropy appears to be small, starting from $\sim 1$ close to $T_c(H = 0)$ and increasing to $\sim 1.2$ at $T = 1/2~T_c(H = 0)$. This $H_{c2}$ anisotropy is smaller than $\gamma \leq 6$ reported for NdFeAs(O$_{1-x}$F$_x$), \cite{jia08a} and somewhat smaller, but similar to, the $\gamma \leq 2$ found in K- or Co- doped, superconducting BaFe$_2$As$_2$ crystals. \cite{nin08b,alt08a,yua09a,nin08c}. The $H_{c2}$ values in CaFe$_2$As$_2$ under pressure are by approximately factor of 2 lower than that in the Ba(Fe$_{1-x}$Co$_x$)$_2$As$_2$ samples with similar $T_c(H=0)$ at ambient pressure, \cite{nin08c} these difference may be due to the larger impurity scattering (i.e. shorter mean free path) in the superconducting members of the Ba(Fe$_{1-x}$Co$_x$)$_2$As$_2$ series.
\\

\begin{figure}
\begin{center}
\includegraphics[angle=0,width=90mm]{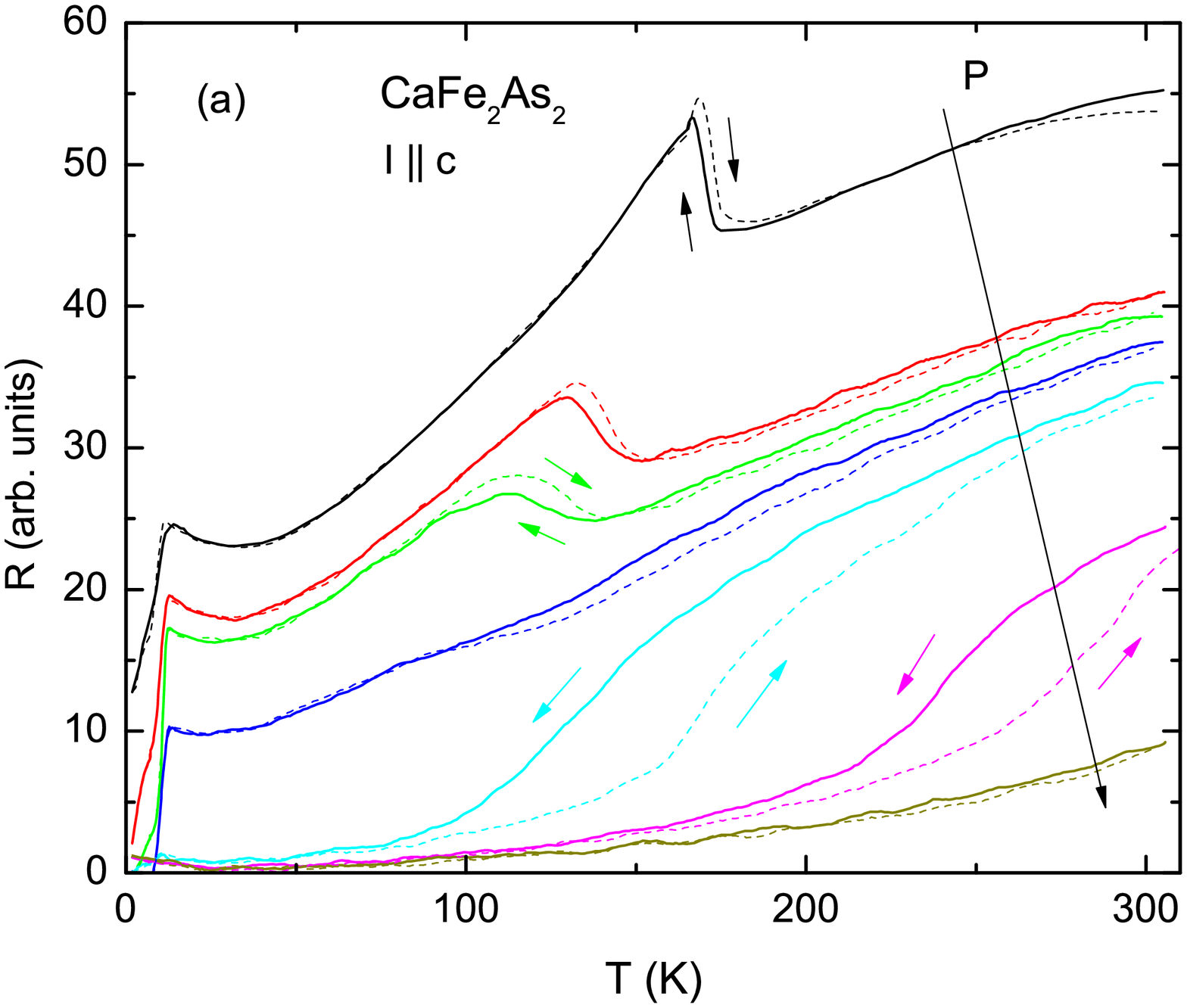}
\includegraphics[angle=0,width=90mm]{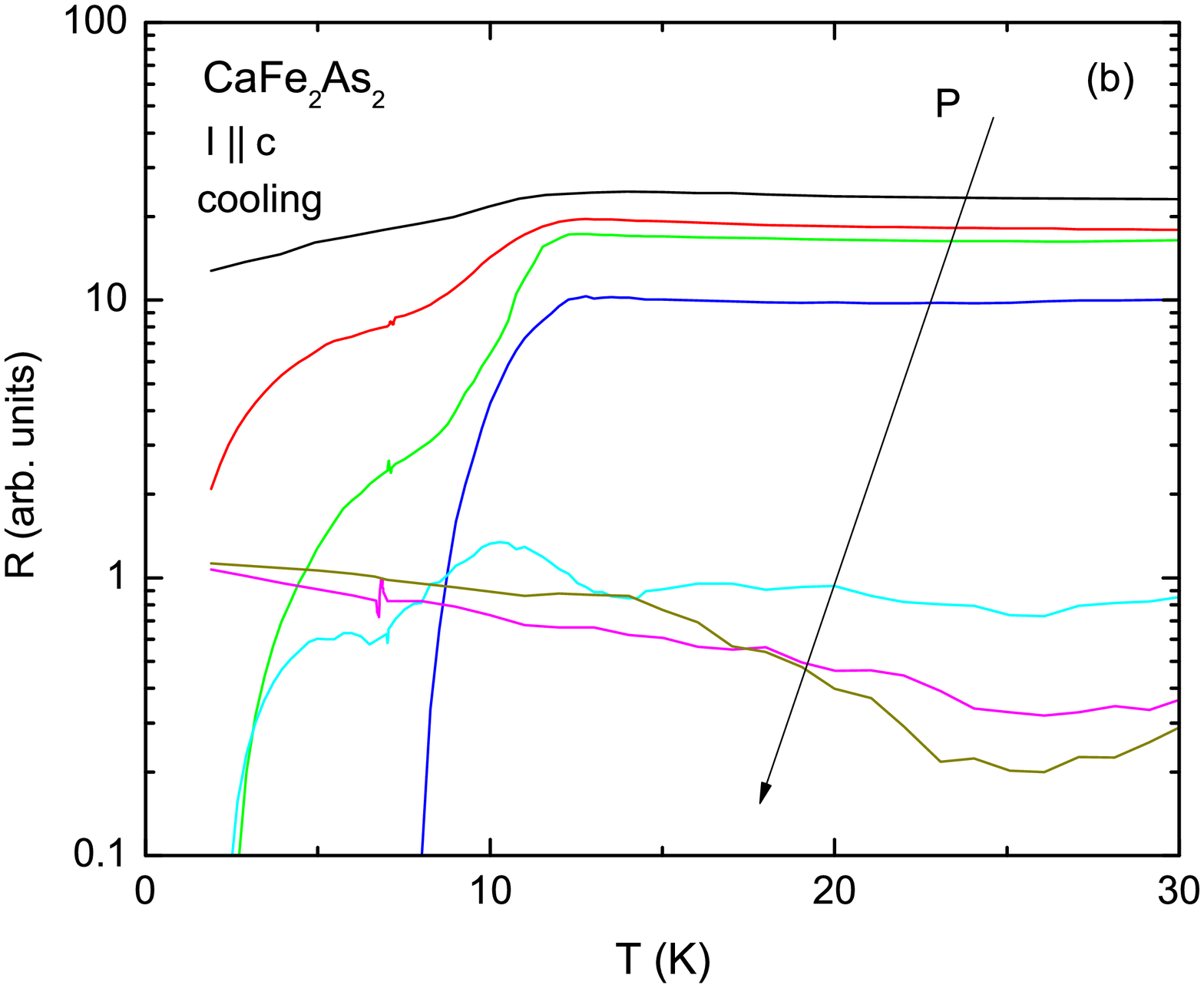}
\end{center}
\caption{(Color online) (a) Temperature-dependent resistance for $I \| c$ (see the text) of CaFe$_2$As$_2$ at different applied pressures (0, 2.0, 3.1, 4.3, 5.8, 12.0, and 17.4 kbar, arrow is pointing to the direction of increase of pressure). Data taken on cooling are shown as solid lines, that taken on warming are dashed lines.  Panel (b) - low temperature part of the data in the panel (a) with only the data taken on cooling shown.}\label{F7}
\end{figure}

The temperature-dependent resistance curves at different pressures measured with the current flowing approximately along the $c$-crystallographic axis (both on cooling and on warming) are shown in Fig. \ref{F7}(a). Taking into account the morphology of the CaFe$_2$As$_2$ crystals, possible errors on dimensions and the geometry of the contacts for these measurements, we can (conservatively) state that the anisotropy of resistivity in CaFe$_2$As$_2$ at ambient pressure is close to unity (within the factor of 2). \cite{tan09a} The structural/antiferromagnetic phase transition at ambient pressure is marked by an anomaly similar in form to that in in-plane resistivity, $\rho_{ab}$. This anomaly moves down and broadens under pressure but remains, at comparable pressures, somewhat sharper than that seen for $\rho_{ab}$. At higher pressures, the tetragonal to collapsed tetragonal phase transition is seen in $\rho_c(T)$ similarly to the way it is observed in $\rho_{ab}(T)$. The hysteresis of the T - cT phase transition is again significantly larger than that for tetragonal - orthorhombic/AFM transition, consistent with earlier observations. \cite{tor08a,lee08a} A shallow minimum in the $c$-axis resistivity is seen at $20 - 30$ K for all pressures studied. Finally, a sharp drop in $\rho_c(T)$ at $T \sim 10$ K is seen in $P = 0$ and $P = 2.0$ kbar data (Fig. \ref{F7}(b)) possibly pointing to some incomplete, filamentary superconductivity. For the next pressure, $P = 3.1$ kbar, a complete, with $\rho = 0$, superconducting transition (although with a low-temperature knee) is observed. At the next pressure (4.3 kbar) this transition is sharp and without a knee, then it broadens again at 5.8 kbar and is not seen at all at 12 kbar and higher.

\begin{figure}
\begin{center}
\includegraphics[angle=0,width=120mm]{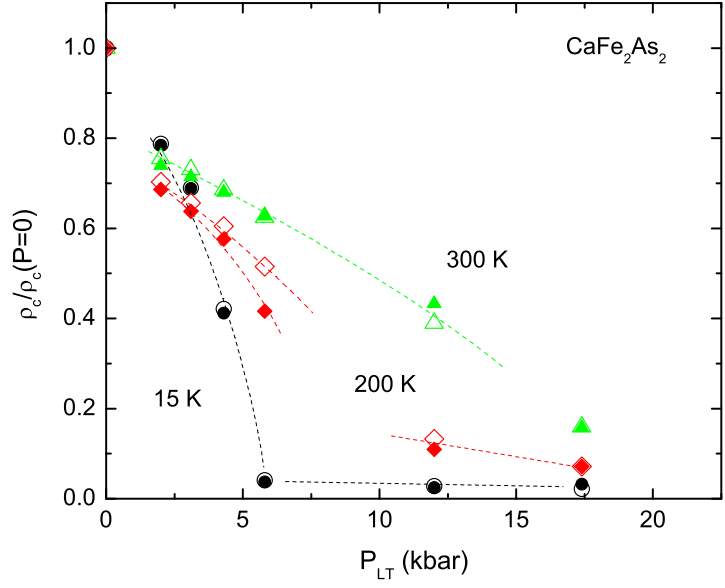}
\end{center}
\caption{(Color online) Normalized to the values at $P = 0$ room temperature resistivity, $\rho_c (300K)$, resistivity at 200 K, $\rho_c (200K)$, and low temperature - normal state resistivity, $\rho_c (15K)$ for CaFe$_2$As$_2$ under pressure. Open symbols - data taken on warming, solid symbols - data taken on cooling.}\label{F8}
\end{figure}

\begin{figure}
\begin{center}
\includegraphics[angle=0,width=120mm]{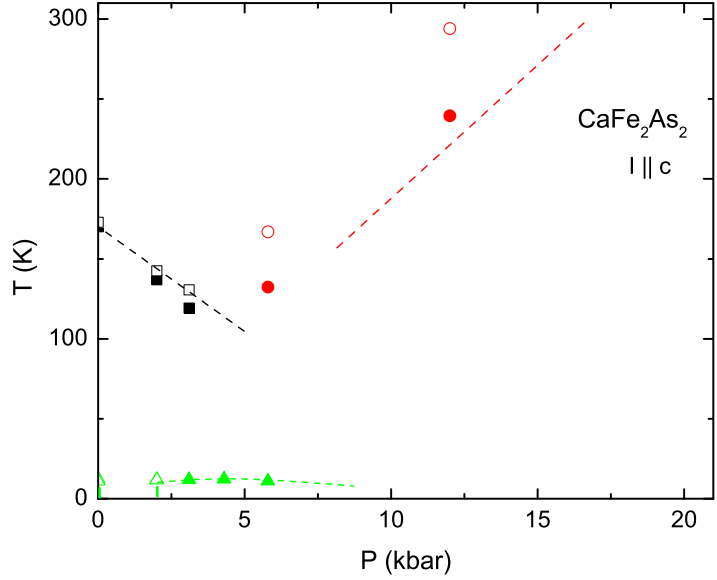}
\end{center}
\caption{(Color online) Pressure-temperature phase diagram of CaFe$_2$As$_2$ as obtained from temperature-dependent resistance for $I \| c$ (see text) measurements at different pressures. Solid triangles - onset $T_c$ for complete superconducting transitions; open triangles - for incomplete superconducting transitions; squares and circles: tetragonal - orthorhombic/AFM and tetragonal - collapsed tetragonal phase transitions (respectively), solid symbols - on cooling, open - on warming. Dashed lines sketch phase boundaries from $I \| ab$ data in Fig. \ref{F3}.}\label{F9}
\end{figure}

The normal state, $c$-axis resistivity changes significantly under pressure (Fig. \ref{F8}). The room temperature resistivity decreases continuously under pressure and by $\sim 17$ kbar becomes approximately factor of 6 smaller than at ambient pressure. For 3 kbar $\leq P \leq$ 6 kbar the resistivity at 15 K decreases rapidly, down to $\sim 4\%$ of its  $P = 0$ value and then, for higher pressures, practically does not change. This remarkable decrease in $\rho_c$ is the primary reason the high pressure, low temperature resistivity data (Fig. \ref{F7}(b)) appears to be so noisy (on a semi-log plot). As was the case for $a$-axis resistivity, this rapid drop helps define the pressure range over which superconductivity is detected. Qualitatively, the behavior of the room temperature and low temperature normal state $\rho_{ab}$ and $\rho_c$ resistivities under pressure is very similar, but with the relative changes in $\rho_c$ being larger. It should be noted, that pressure may be promoting better connectivity between the sample layers and by this contribute to larger relative effect of pressure on $\rho_c$. As expected, the $\rho_c (T)$ data yield a $T - P$ phase diagram (Fig. \ref{F9}) very similar to that obtained from $\rho_{ab}$ measurements (Ref. \onlinecite{tor08a} and Fig. \ref{F3}) (for comparison the phase lines inferred from data shown in Fig. \ref{F3} are also shown).
\\

\begin{figure}
\begin{center}
\includegraphics[angle=0,width=120mm]{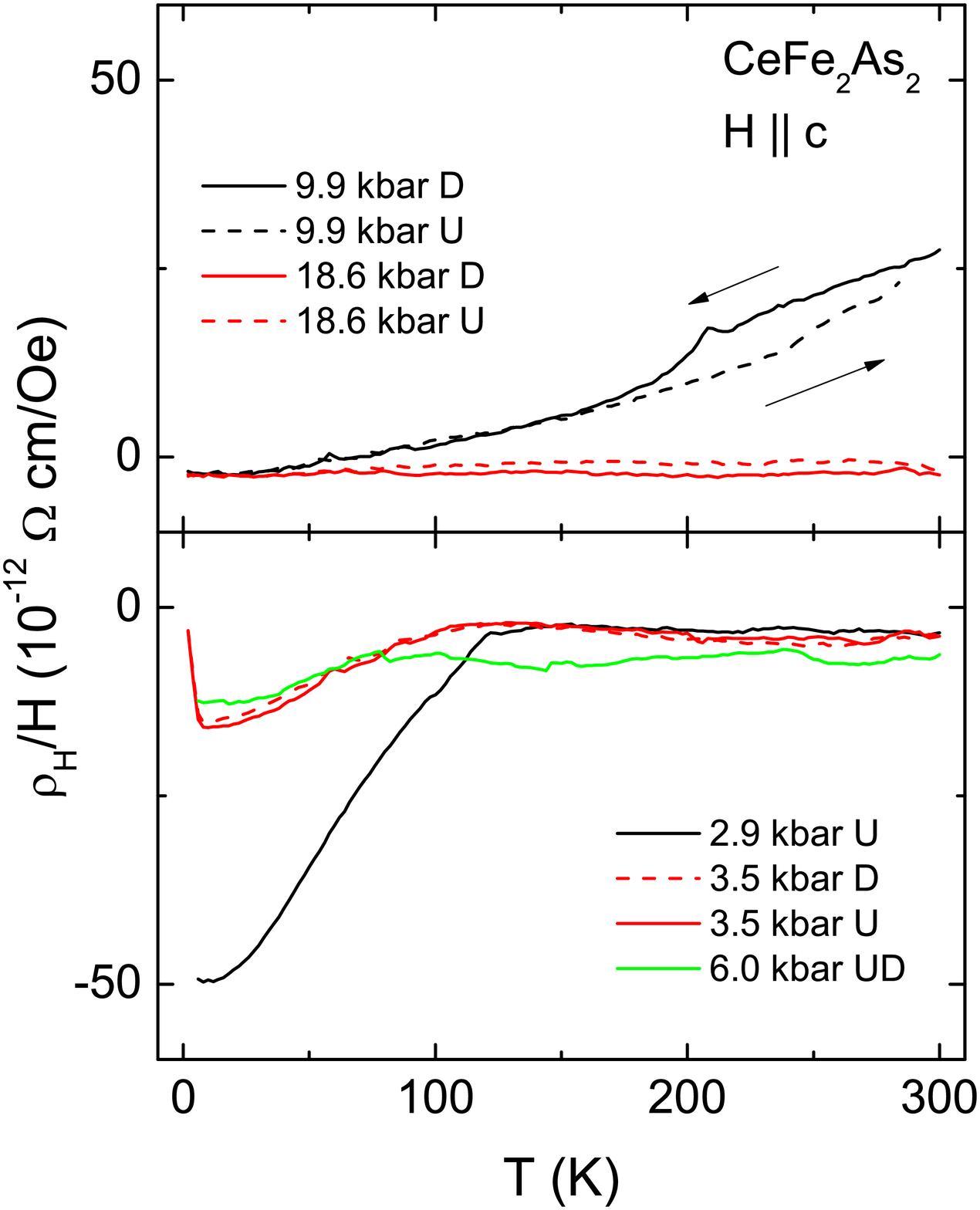}
\end{center}
\caption{(Color online) Temperature-dependent Hall coefficient $\rho_H/H$ measured at different pressures. Measurements on cooling and warming are marked "D" and "U" respectively in the legend.}\label{F10}
\end{figure}

The temperature-dependent Hall coefficient, $\rho_H/H$ (where $\rho_H$ is the Hall resistance), measured for different applied pressures, is plotted in Fig. \ref{F10}. As pressure is increased ($P = 2.9$ and 3.5 kbar), the low temperature downturn associated with orthorhombic/AFM phase becomes smaller and shifts to lower temperatures, consistent with the $T - P$ phase diagram. For the 3.5 kbar curves, superconductivity (in the 90 kOe applied field used for the measurements) is signaled by the sharp low temperature upturn (towards $\rho_H/H = 0$) with the onset at $\sim 6$ K, grossly consistent with the $H_{c2}(T)$ data at different pressures in Fig. \ref{F5}. The 9.9 kbar data show a signature associated with T - cT transition that is broad and hysteretic, again consistent with the resistivity data (see Fig. \ref{F2}(b) for a comparable data set). It should be noted that the overall temperature behavior of 9.9 kbar curves is different from other sets of data, however one should have in mind that this is the only pressure at which the T - cT line is clearly crossed below room temperature.  Finally, at 18.6 kbar the Hall coefficient is small and featureless, as one can expect for non-magnetic, normal metal.
\\

\begin{figure}
\begin{center}
\includegraphics[angle=0,width=120mm]{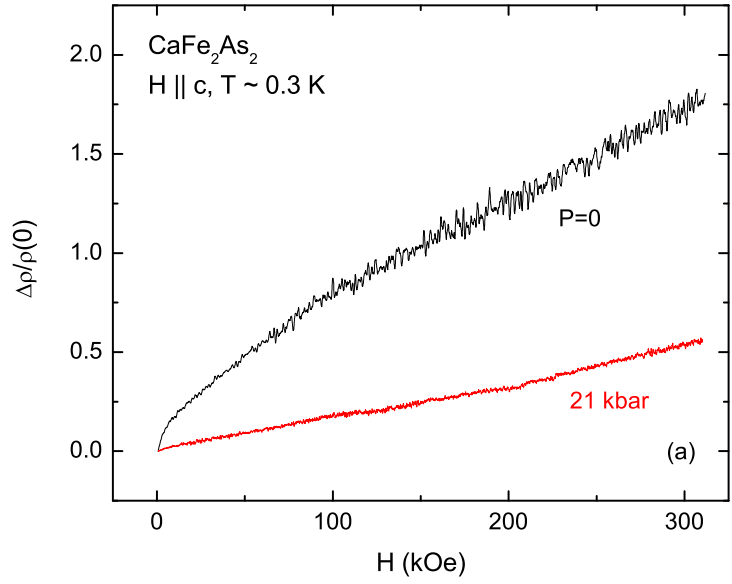}
\end{center}
\caption{(Color online) Normalized low temperature ($T \approx 0.3$ K) magnetoresistivity at $P = 0$ and $P \approx 21$ kbar for  CaFe$_2$As$_2$. $I \| ab$, $H \| c$.}\label{F11}
\end{figure}

The normal state, low temperature ($T \approx 0.3$ K) magnetoresistivity (MR) was measured at ambient pressure (in the orthorhombic/AFM phase) and $P \approx 21$ kbar (in the collapsed tetragonal phase) with the current in the $ab$ plane and magnetic field applied along the $c$-axis (Fig. \ref{F11}). Whereas no Shubnikov - de Haas oscillations were observed, the magnetoresistivity is rather high. The overall behavior changes from sub-linear at ambient pressure to very close to linear at 21 kbar. Naively, one would expect that, since (i) the low temperature normal state resistivity decreases by almost an order of magnitude under pressure and (ii) there is no magnetism (often considered as giving a negative contribution to the MR) in the cT phase, the normalized MR would be {\it higher} in the cT phase (e.g. by invoking Kohler's rule, \cite{aaabook,bud98a,mye99a} $\Delta \rho /\rho_0 = F(H/\rho_0)$,  where $\Delta \rho = \rho(T,H) - \rho(T,0)$ and $\rho_0 = \rho(T,0)$). In the experiment, it is more than a factor of 2 {\it lower} and its field dependence is greatly reduced and very close to linear. Linear MR itself is not a very common phenomenon. It was observed experimentally in some anisotropic semimetals (e.g. Refs. [\onlinecite{bud98a,mye99a}]), attracted considerable theoretical attention \cite{aaa} and is still a subject of discussions. \cite{huj08a} Clearly, more effort is required to address the origin of linear MR under pressure in  CaFe$_2$As$_2$. The significant change of low temperature MR in these materials under pressure brings attention to potentially interesting normal state properties of CaFe$_2$As$_2$ and related Fe-As based materials.
\\

\begin{figure}
\begin{center}
\includegraphics[angle=0,width=120mm]{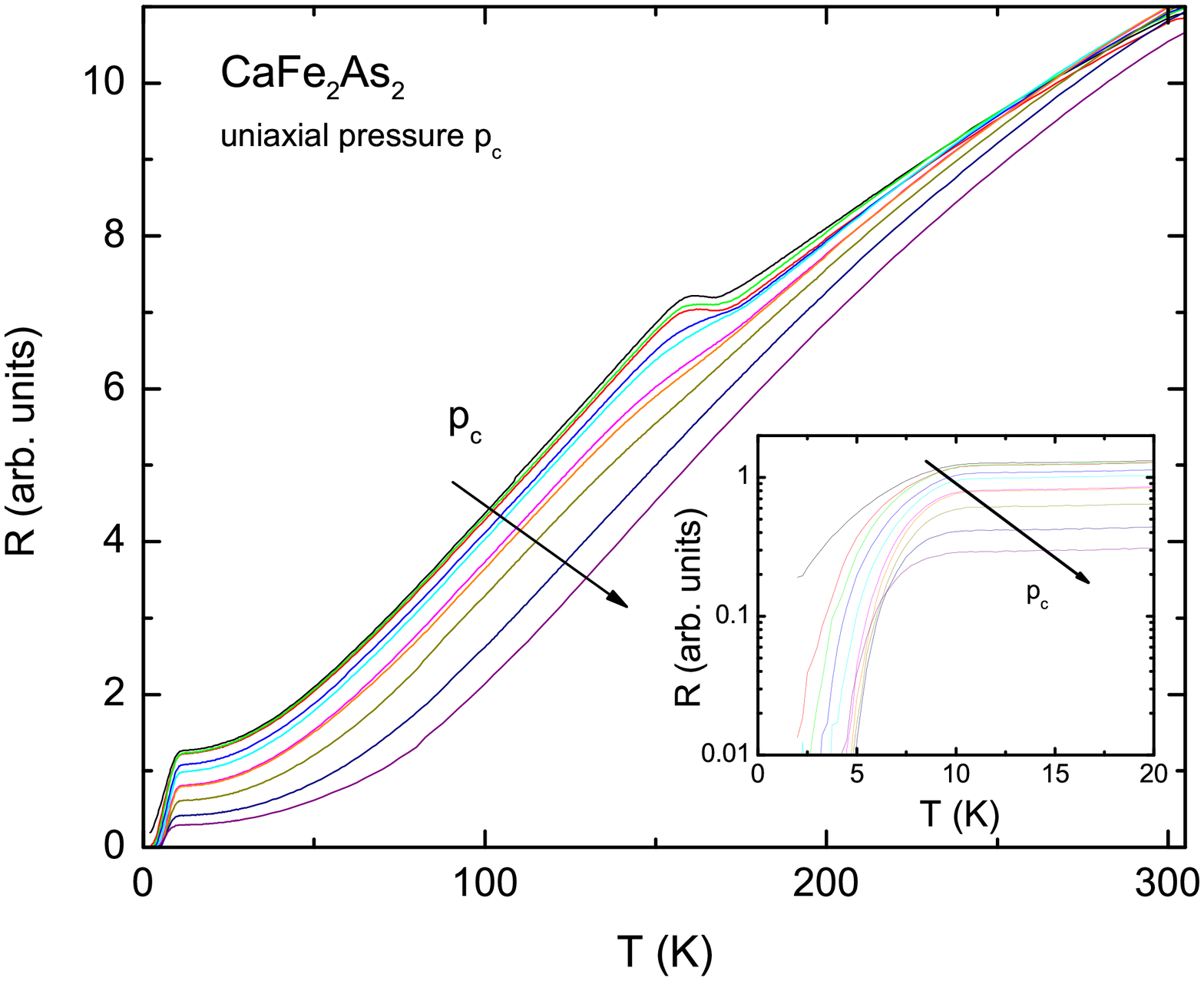}
\end{center}
\caption{(Color online) Temperature-dependent in-plane resistance of CaFe$_2$As$_2$ at different applied uniaxial pressure (0.5, 0.7, 0.8, 1.1, 1.4, 1.7, 2.1, 2.4, 2.7, 2.8 kbar, arrow is pointing to the direction of increase of pressure). Data taken on cooling are shown. Inset: low temperature part of the data.}\label{F12}
\end{figure}

\begin{figure}
\begin{center}
\includegraphics[angle=0,width=90mm]{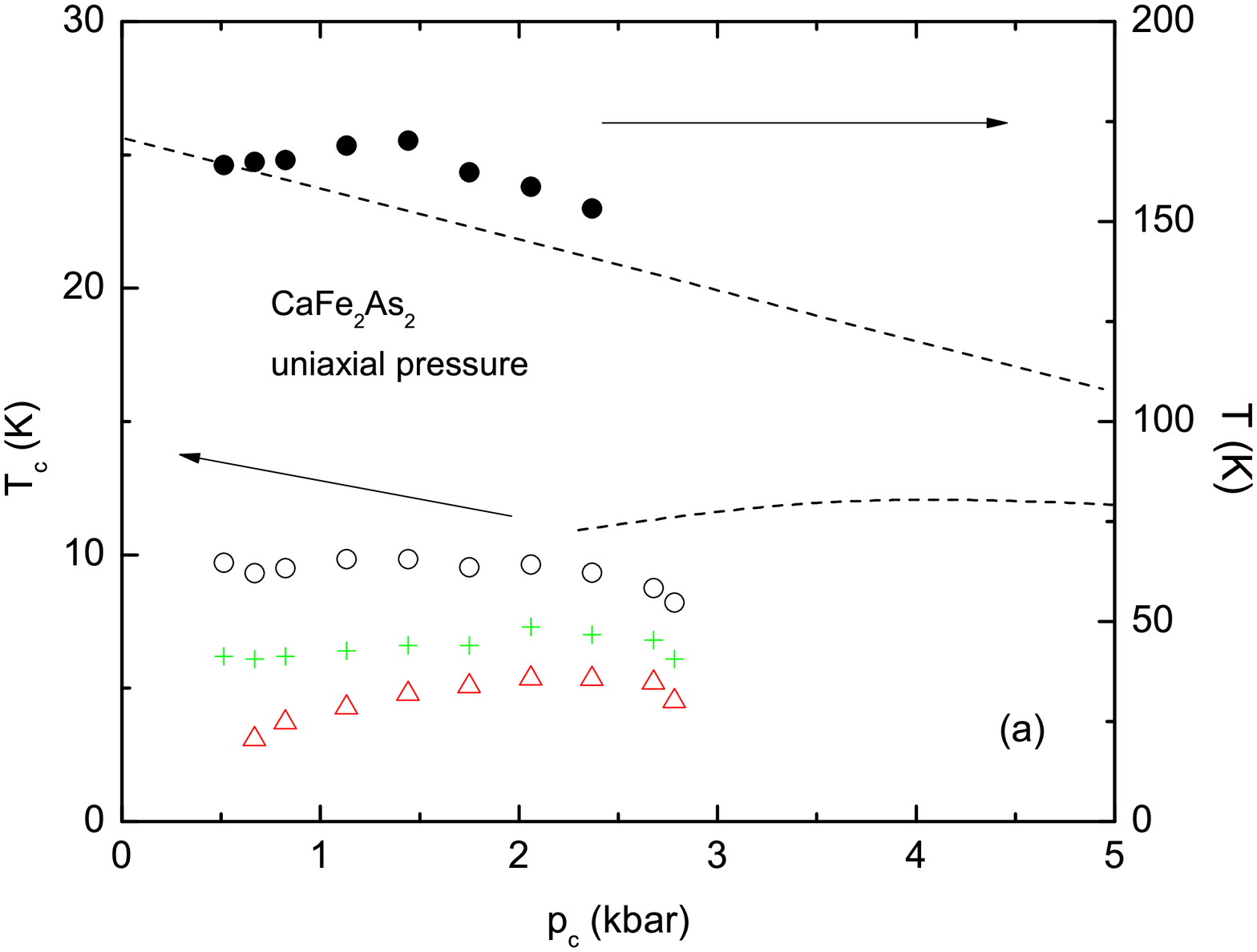}
\includegraphics[angle=0,width=90mm]{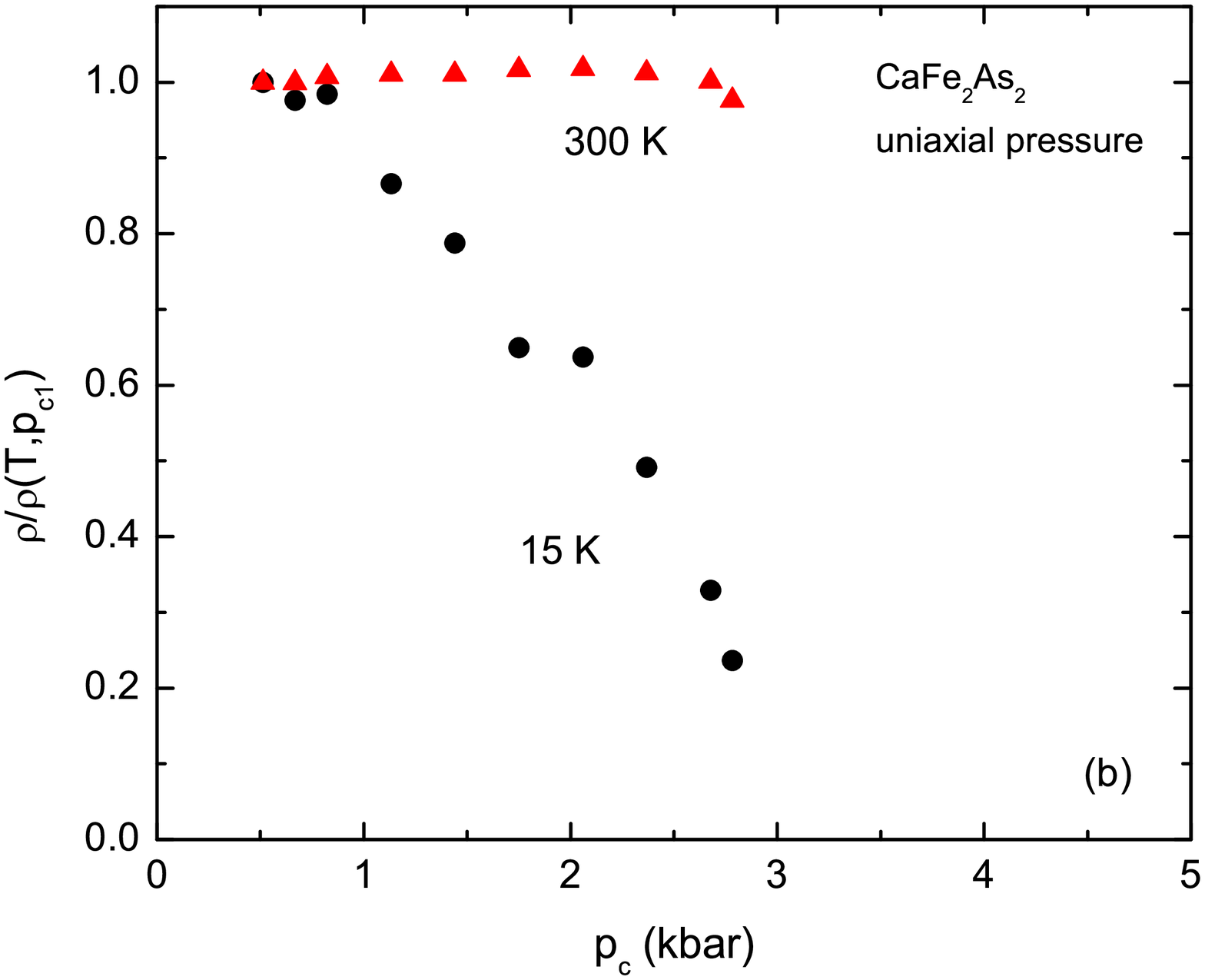}
\end{center}
\caption{(Color online) (a) Superconducting transition temperature (left axis) and structural/AMF transition temperature (right axis) as a function of uniaxial pressure. Symbols: circles - $T_c$ onset, triangles - $T_c$ offset, crosses - maximum in $dR/dT$ derivatives; structural/AMF transition temperature was defined as a beginning of upturn in resistance (note that an alternative criterion, minimum in $dR/dT$ gives similar results). Dashed lines - sketches of the phase lines from liquid medium cell measurements (Fig. \ref{F3}). (b) Resistivity as a function of applied, uniaxial pressure for $T = 300$ K and $T = 15$ K. Each curve is normalized to the lowest pressure value of the resistivity.}\label{F13}
\end{figure}

Given that non-hydrostatic components of pressure may be key to stabilizing superconductivity in CaFe$_2$As$_2$, preliminary measurements of in-plane resistance under uniaxial pressure applied along the $c$-axis were performed (Fig. \ref{F12}). A complete ($R = 0$) superconducting transition in resistance is observed already at $p_c \approx 0.7$ kbar,  however this transition is rather wide. The high temperature feature associated with the structural/AFM transition at the ambient pressure broadens significantly under uniaxial pressure, with a gradual suppression of its position (Fig. \ref{F13}(a)). In the uniaxial pressure range of this study, the room temperature (300 K) resistivity decreases only modestly at the highest pressures, whereas low temperature, normal state (15 K) resistivity decreases by almost a factor of five as uniaxial pressure is increased (Fig. \ref{F13}(b)). The overall behavior under uniaxial stress, $p_c$ is very similar to that under pressure in clamp piston-cylinder cell using liquid media. The slight difference is that a complete superconducting transition in resistivity is observed at significantly lower pressure leading to what appears to be a wide range of $p_c$ with apparent coexistence of superconductivity and magnetism. In order to make the comparison between our uniaxial and liquid medium cell $T_c(P)$ data the phase lines from Fig. \ref{F3} have been added to Fig. \ref{F13}(a). Within our limited range of $p_c$ the agreement is noteworthy. It is possible that the low $p_c$ limit ($p_c < 3$ kbar) in our measurements prevented the observation of the T - cT transition under uniaxial pressure, although, as will be discussed below, it is also possible that the cT phase does not appear under uniaxial pressure.

\begin{figure}
\begin{center}
\includegraphics[angle=0,width=120mm]{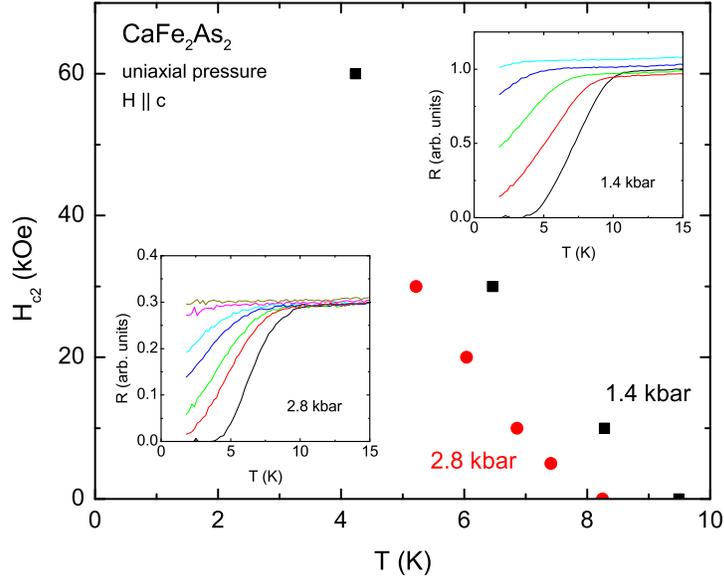}
\end{center}
\caption{(Color online) Temperature-dependent upper critical field ($H_{c2}$) determined at 1.4 kbar and 2.8 kbar uniaxial pressures from the onsets of resistive transitions. Insets: low temperature resistivities measured in different magnetic fields (0, 10, 30, 60, 90 kOe for $p_c = 1.4$ kbar and 0, 5, 10, 20, 30, 60, 90 kOe for $p_c = 2.8$ kbar) applied along the $c$-axis.}\label{F14}
\end{figure}

The superconducting transition,  as seen in resistance, broadens significantly in applied field to the extent of becoming incomplete at $\sim 10$ kOe (Fig. \ref{F14}, insets). This appears to be consistent with a small, just allowing for percolation in zero field, superconducting fraction. If, nevertheless,  the onset of the transition is used to determine $H_{c2}(T)$(Fig. \ref{F14}), the results are similar to that obtained under liquid medium ("hydrostatic") pressure.

\section{Summary}

\begin{figure}
\begin{center}
\includegraphics[angle=0,width=120mm]{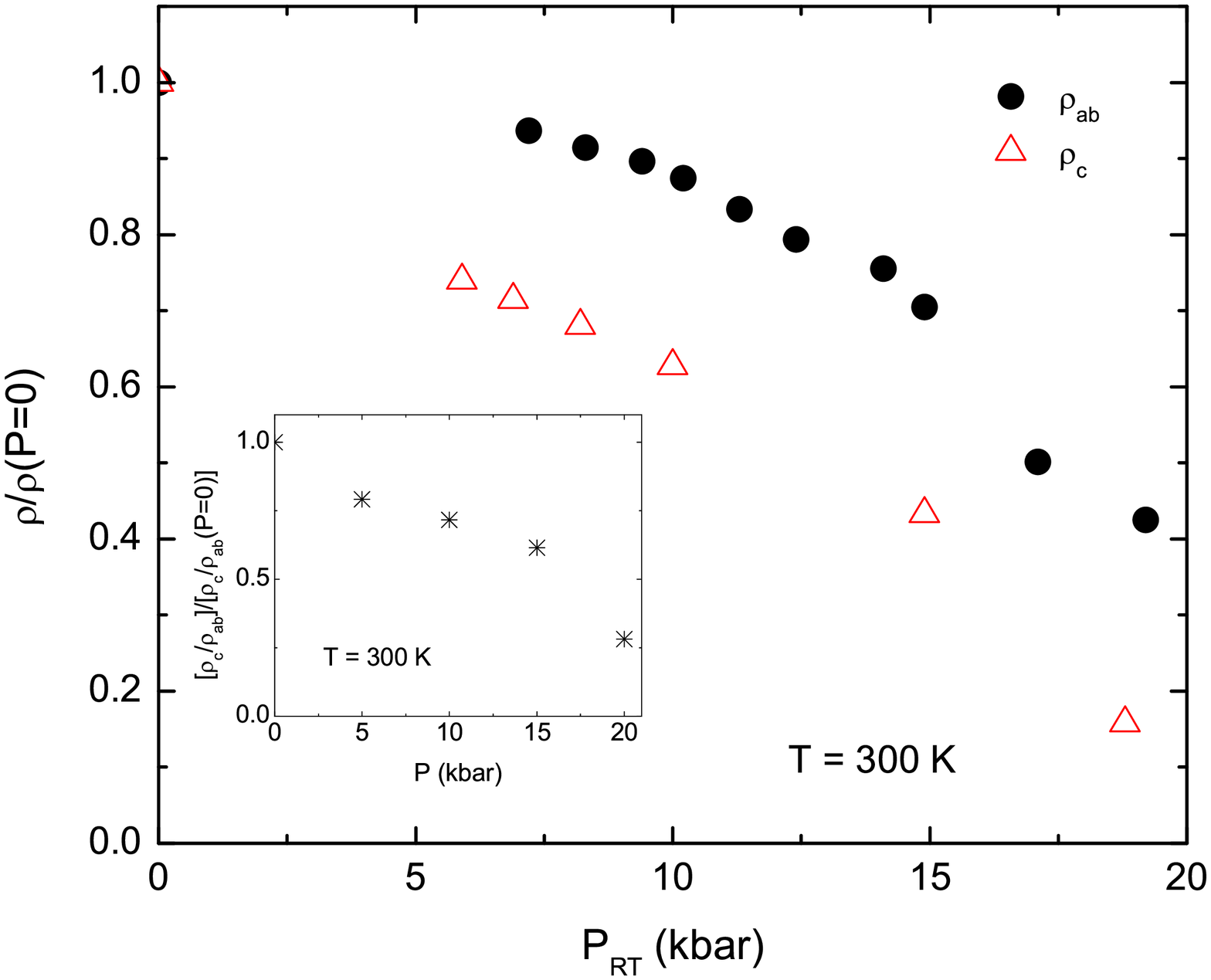}
\end{center}
\caption{(Color online) Normalized to the values at $P = 0$ room temperature resistivities, $\rho_{ab} (300K)$ and $\rho_c (300K)$ as a function of room temperature values of pressure. Inset: estimate of the relative change of the room temperature resistivity anisotropy under pressure.}  \label{F15}
\end{figure}

The results of this work show that the $P - T$ phase diagram of CaFe$_2$As$_2$ is robust and reproducible when measured in a liquid media and that the stabilized superconductivity manifests a similar, small, $H_{c2}(T)$ anisotropy as the other AEFe2As2 materials. \cite{alt08a,yua09a,nin08c} Phase diagrams assembled from resistivity measurements with current flowing along the $ab$ plane, or along the $c$-axis are essentially identical and even when the $P - T$ phase diagram is determined using $c$-axis, uniaxial pressure there is good agreement, at least over the limited uniaxial pressure range available to us. It is noteworthy that the $c$-axis resistivity at room temperature appear to decrease faster than the in-plane resistivity (Fig. \ref{F15}). If we assume the ambient pressure resistivity anisotropy of approximately two found in Ref. \onlinecite{tan09a}, then the resistivity anisotropy decreases under pressure, making the material even more three-dimensional.

Qualitatively, the evolution of Hall resistivity in CaFe$_2$As$_2$ under pressure is similar to that in BaFe$_2$As$_2$ with Co-doping \cite{fan09a,rul09a,mun09a} in the lower pressure part of the $P - T$ phase diagram (suppression of the structural/AFM transition and observation of superconductivity). The current Hall data set, in the region of and above the T - cT transition, is somewhat sparse, and may be affected by the multi-crystallographic state of the sample at intermediate pressures  and clearly requires further, detailed studies.

An apparent violation of the Kohler's rule in magnetoresistance under pressure for CaFe$_2$As$_2$ points out to significant difference in the band structure and/or scattering between the low temperature orthorhombic/AFM (at ambient pressure) and cT phase (at 19 kbar) that makes the Kohler's rule unapplicable.

For the uniaxial, $c$-axis, pressure measurements the existence of the T - cT phase line remains an open question. Either, given our relatively low, maximum uniaxial pressure we are just short of reaching the cT transition, or, bearing in mind the drop in the low temperature, residual, resistance, the uniaxial pressure could allow the system to bypass the cT phase by gradually changing the $c$-axis and removing the driving force for the cT transition.  The further uniaxial measurements should clarify this ambiguity.

\begin{acknowledgments}
Work at the Ames Laboratory was supported by the US Department of Energy - Basic Energy Sciences under Contract No. DE-AC02-07CH11358.  MST gratefully acknowledges support of the National Science Foundation under DMR-0306165 and DMR-0805335. A portion of this work was performed at the National High Magnetic Field Laboratory, which is supported by NSF Cooperative Agreement No. DMR-0084173, by the State of Florida, and by the US DOE.
\end{acknowledgments}

\clearpage

\clearpage

\clearpage

\clearpage

\clearpage

\clearpage

\clearpage

\clearpage

\clearpage

\clearpage

\clearpage

\clearpage

\clearpage

\clearpage

\begin{thebibliography}{99}

%\bibitem{nor08a}  Michael R. Norman, Physics  {\bf 1}, 21 (2008) (\verb=http://physics.aps.org/articles/v1/21=).

\bibitem{nin08a} N. Ni, S. Nandi, A. Kreyssig, A. I. Goldman, E. D. Mun, S. L. Bud'ko, and P. C. Canfield, Phys. Rev. B {\bf 78},  014523 (2008).

\bibitem{ron08a} F. Ronning, T. Klimczuk, E. D. Bauer, H. Volz, and J. D. Thompson, J. Phys.: Condens. Matter {\bf 20}, 322201 (2008).

\bibitem{wug08a} G. Wu, H. Chen, T. Wu, Y. L. Xie, Y. J. Yan, R. H. Liu, X. F. Wang, J. J. Ying and X. H. Chen, J. Phys.: Condens. Matter {\bf 20}, 422201 (2008).
	
\bibitem{gol08a} A. I. Goldman, D. N. Argyriou, B. Ouladdiaf, T. Chatterji, A. Kreyssig, S. Nandi, N. Ni, S. L. Bud'ko, P. C. Canfield, and R. J. McQueeney, Phys. Rev. B {\bf 78}, 100506 (2008).	

\bibitem{rot08a} Marianne Rotter, Marcus Tegel, Inga Schellenberg, Wilfried Hermes, Rainer P\"{o}ttgen, Dirk Johrendt, Phys. Rev. B {\bf 78}, 020503 (2008)

\bibitem{nin08b} N. Ni, S. L. Bud'ko, A. Kreyssig, S. Nandi, G. E. Rustan, A. I. Goldman, S. Gupta, J. D. Corbett, A. Kracher, and P. C. Canfield Phys. Rev. B {\bf 78}, 014507 (2008).

\bibitem{mit08a} R. Mittal, Y. Su, S. Rols, T. Chatterji, S. L. Chaplot, H. Schober, M. Rotter, D. Johrendt, and Th. Brueckel, Phys. Rev. B {\bf 78}, 104514 (2008).

\bibitem{kre08a} C. Krellner, N. Caroca-Canales, A. Jesche, H. Rosner, A. Ormeci, and C. Geibel, Phys. Rev. B {\bf 78}, 100504 (2008).

\bibitem{yan08a} J.-Q. Yan, A. Kreyssig, S. Nandi, N. Ni, S. L. Bud'ko, A. Kracher, R. J. McQueeney, R. W. McCallum, T. A. Lograsso, A. I. Goldman, and P. C. Canfield, Phys. Rev. B {\bf 78}, 024516 (2008).

\bibitem{zha08a} Jun Zhao, W. Ratcliff, J. W. Lynn, G. F. Chen, J. L. Luo, N. L. Wang, Jiangping Hu, and Pengcheng Dai, Phys. Rev. B {\bf 78}, 140504 (2008).

\bibitem{mcq08a} R. J. McQueeney, S. O. Diallo, V. P. Antropov, G. D. Samolyuk, C. Broholm, N. Ni, S. Nandi, M. Yethiraj, J. L. Zarestky, J. J. Pulikkotil, A. Kreyssig, M. D. Lumsden, B. N. Harmon, P. C. Canfield, and A. I. Goldman, Phys. Rev. Lett. {\bf 101}, 227205 (2008).
    	
\bibitem{tor08a} Milton S. Torikachvili, Sergey L. Bud'ko, Ni Ni, and Paul C. Canfield, Phys. Rev. Lett. {\bf 101}, 057006 (2008).

\bibitem{par08a} Tuson Park, Eunsung Park, Hanoh Lee, T. Klimczuk, E. D. Bauer, F. Ronning and J. D. Thompson, J. Phys.: Condens. Matter {\bf 20},  322204 (2008).

\bibitem{ali08a} Patricia L. Alireza, Y. T. Chris Ko, Jack Gillett, Chiara M. Petrone, Jacqueline M. Cole, Gilbert G. Lonzarich and Suchitra E. Sebastian, J. Phys.: Condens. Matter {\bf 21}, 012208 (2009).

\bibitem{iga08a} K. Igawa, H. Okada, H. Takahashi, S. Matsuishi, Y. Kamihara, M. Hirano, H. Hosono, K. Matsubayashi, Y. Uwatoko, J. Phys. Soc. Jpn. {\bf 78}, 025001 (2009).

\bibitem{kot08a} Hisashi Kotegawa, Hitoshi Sugawara, and Hideki Tou, J. Phys. Soc. Jpn. {\bf 78} 013709 (2009).

\bibitem{man09a} Awadhesh Mani, Nilotpal Ghosh, S. Paulraj, A. Bharathi, and C. S. Sundar, arXiv:0903.4236v1, unpublished.

\bibitem{col09a} E. Colombier, S. L. Bud'ko, N. Ni, and P. C. Canfield, arXiv:0904.4488v1, unpublished.

\bibitem{kot09a} Hisashi Kotegawa, Takayuki Kawazoe, Hitoshi Sugawara, Keizo Murata, and Hideki Tou1, arXiv:0904.4631v1, unpublished.

\bibitem{kre08b} A. Kreyssig, M. A. Green, Y. Lee, G. D. Samolyuk, P. Zajdel, J. W. Lynn, S. L. Bud'ko, M. S. Torikachvili, N. Ni, S. Nandi, J. Le\~{a}o, S. J. Poulton, D. N. Argyriou, B. N. Harmon, P. C. Canfield, R. J. McQueeney, A. I. Goldman, Phys. Rev. B {\bf 78}, 184517 (2008).

\bibitem{gok08a} T. Goko, A. A. Aczel, E. Baggio-Saitovitch, S. L. Bud'ko, P.C. Canfield, J. P. Carlo, G. F. Chen, Pengcheng Dai, A. C. Hamann, W. Z. Hu, H. Kageyama, G. M. Luke, J. L. Luo, B. Nachumi, N. Ni, D. Reznik, D. R. Sanchez-Candela, A. T. Savici, K. J. Sikes, N. L. Wang, C. R. Wiebe, T. J. Williams, T. Yamamoto, W. Yu, and Y. J. Uemura, arXiv:0808.1425v1, unpublished.

\bibitem{lee08a} Hanoh Lee, Eunsung Park, Tuson Park, F. Ronning, E. D. Bauer, and J. D. Thompson, arXiv:0809.3550v1, unpublished.

\bibitem{yuw08a} W. Yu, A. A. Aczel, T. J. Williams, S. L. Bud'ko, N. Ni, P. C. Canfield, and G. M. Luke, Phys. Rev. B {\bf 79}, 020511 (2009).

\bibitem{gol08b} A. I. Goldman, A. Kreyssig, K. Proke\u{s}, D. K. Pratt, D. N. Argyriou, J. W. Lynn, S. Nandi, S. A. J. Kimber, Y. Chen, Y. B. Lee, G. Samolyuk, J. B. Le\~{a}o, S. J. Poulton, S. L. Bud'ko, N. Ni, P. C. Canfield, B. N. Harmon, and R. J. McQueeney, Phys. Rev. B {\bf 79}, 024513 (2009).

\bibitem{lee08b} Hanoh Lee, Eunsung Park, Tuson Park, F. Ronning, E. D. Bauer, and J. D. Thompson, arXiv:0809.3550v2, unpublished.

\bibitem{yil08a} T. Yildirim, Phys. Rev. Lett. {\bf 102}, 037003 (2009).

\bibitem{sam08a} G. D. Samolyuk, and V. P. Antropov, Phys. Rev. B {\bf 79}, 052505 (2009).

\bibitem{can09a} P. C. Canfield, S. L. Bud'ko, N. Ni, A. Kreyssig, A. I. Goldman, R. J. McQueeney, M. S. Torikachvili, D. N. Argyriou, G. Luke, and W. Yu, Physica C, {\bf 469} 404 (2009).

\bibitem{can92a} P. C. Canfield, and Z. Fisk, Philos. Mag. B {\bf 65}, 1117 (1992).

\bibitem{eil81a} A. Eiling, and J. S. Schilling, J. Phys. F {\bf 11}, 623 (1981).

\bibitem{son03a} C. Song, Jaehyun Park, Japil Koo, K.-B. Lee, J. Y. Rhee, S. L. Bud'ko, P. C. Canfield, B. N. Harmon, and A. I. Goldman, Phys. Rev. B {\bf 68}, 035113 (2003).

\bibitem{led} H. M. Ledbetter, J. Appl. Phys., {\bf 52} 1587 (1981); Hassel Ledbetter, Mat. Sci. Eng. A {\bf 442} 31 (2006) and references therein.

\bibitem{jia08a} Ying Jia, Peng Cheng, Lei Fang, Huiqian Luo, Huan Yang, Cong Ren, Lei Shan, Changzhi Gu, and Hai-Hu Wen, Appl. Phys. Lett., {\bf 93}, 032503 (2008); Ying Jia, Peng Cheng, Lei Fang, Huan Yang, Cong Ren, Lei Shan, Chang-Zhi Gu, and Hai-Hu Wen, Supercond. Sci. Tech., {\bf 21}, 105018 (2008).

\bibitem{alt08a} M. M. Altarawneh, K. Collar, C. H. Mielke, N. Ni, S. L. Bud'ko, and P. C. Canfield, Phys. Rev. B {\bf 78}, 220505 (2008).

\bibitem{yua09a}  H. Q. Yuan, J. Singleton, F. F. Balakirev, S. A. Baily, G. F. Chen, J. L. Luo, N. L. Wang, Nature {\bf 457}, 565 (2009).

\bibitem{nin08c} N. Ni, M. E. Tillman, J.-Q. Yan, A. Kracher, S. T. Hannahs, S. L. Bud'ko, and P. C. Canfield, Phys. Rev. B {\bf 78}, 214515 (2008).

\bibitem{tan09a} M. A. Tanatar, N. Ni, G. D. Samolyuk, S. L. Bud'ko, P. C. Canfield, and R. Prozorov, Phys. Rev. B {\bf 79}, 134528 (2009).

\bibitem{aaabook} see, for example, A. A. Abrikosov, {\it Fundamentals of the Theory of Metals} (North-Holland, Amsterdam) (1988).

\bibitem{bud98a} S. L. Bud'ko, P. C. Canfield, C. H. Mielke, and A. H. Lacerda, Phys. Rev. B {\bf 57}, 13624 (1998), and references therein.

\bibitem{mye99a} K. D. Myers, S. L. Bud'ko, I. R. Fisher, Z. Islam, H. Kleinke, and P. C. Canfield, J. Magn. Magn. Mater. {\bf 205}, 27 (1999).

\bibitem{aaa} A. A. Abrikosov, Phys. Rev. B {\bf 58}, 2788 (1998); A. A. Abrikosov, Europhys. Lett., {\bf 49}, 789 (2000); A. A. Abrikosov, J. Phys. A: Mat. Gen., {\bf 36} 9119 (2003), and references therein.

\bibitem{huj08a} Jingshi Hu, and T. F. Rosenbaum, Nature Materials, {\bf 7}, 697 (2008).

\bibitem{fan09a} Lei Fang, Huiqian Luo, Peng Cheng, Zhaosheng Wang, Ying Jia, Gang Mu, Bing Shen, I. I. Mazin, Lei Shan, Cong Ren, and Hai-Hu Wen,  arXiv:0903.2418v2, unpublished.

\bibitem{rul09a} F. Rullier-Albenque, D. Colson, A. Forget, and H. Alloul, arXiv:0903.5243v1, unpublished.

\bibitem{mun09a} E. D. Mun, S. L. Bud'ko, N. Ni, and P. C. Canfield, unpublished.

\end{thebibliography}
\end{document}